\documentclass{article}
\usepackage{graphicx}
\usepackage{amsmath}

\newtheorem{theorem}{Theorem}
\newtheorem{acknowledgement}[theorem]{Acknowledgement}

\begin{document}

\begin{center}
{\large BERRY CURVATURE IN GRAPHENE: A NEW APPROACH }\\[0pt]
\vskip2cm Pierre Gosselin \\[0pt]
\vskip1cm Institut Fourier, UMR 5582 CNRS-UJF, UFR de Math\'{e}matiques, \\[%
0pt]
Universit\'{e} Grenoble I, \\[0pt]
BP74, 38402 Saint Martin d'H\`{e}res, Cedex,\\[0pt]
France.

\vskip2cm Alain B\'{e}rard, and Herv\'{e} Mohrbach\\[0pt]
\vskip1cm Laboratoire de Physique Mol\'{e}culaire et des Collisions,
ICPMB-FR CNRS 2843,\\[0pt]
Universit\'{e} Paul Verlaine-Metz,\\[0pt]
57078 Metz Cedex 3,\\[0pt]
France.

\vskip 2cm Subir Ghosh\\[0pt]
\vskip 1cm Physics and Applied Mathematics Unit,\\[0pt]
Indian Statistical Institute,\\[0pt]
203 B. T. Road, Calcutta 700108, \\[0pt]
India.
\end{center}

\vskip3cm \textbf{Abstract:}\newline
In the present paper we have directly computed the Berry curvature terms
relevant for Graphene in the presence of an \textit{inhomogeneous} lattice
distortion. We have employed the generalized Foldy Wouthuysen framework,
developed by some of us \cite{ber0,ber1,ber2}. We show that a non-constant
lattice distortion leads to a valley-orbit coupling which is responsible to
a valley-Hall effect. This is similar to the valley-Hall effect induced by
an electric field proposed in \cite{niu2} and is the analogue of the
spin-Hall effect in semiconductors \cite{MURAKAMI, SINOVA}. Our general
expressions for Berry curvature, for the special case of homogeneous
distortion, reduce to the previously obtained results \cite{niu2}. We also
discuss the Berry phase in the quantization of cyclotron motion.


\vskip 1cm \noindent PACS Numbers: \newpage

\section{Introduction}

The latest \textit{avatar} of crystalline carbon - Graphene \cite{rev} - has
evolved into a meeting ground of Particle and Condensed Matter physicists.
The reason is that under certain conditions the electronic excitations in
graphene behave like massless relativistic fermions moving with (energy
independent) Fermi velocity $v_{F}\approx 10^{6}$ $m/s$, that plays the role
of the velocity of light \cite{sem}. The honeycomb lattice of graphene is
made up of two triangular lattices ($A$ and $B$) and the hexagonal Brillouin
zone has two distinct and degenerate Dirac points or valleys ($K_{+}$ and $%
K_{-}$) where the conduction and valence bands meet. In the undoped case the
chemical potential passes through the Dirac points so that the valence band
is filled whereas the conduction band is empty. In the Tight Binding
Approximation, the dynamical equations of motion for the planar electrons
(with spin not taken in to account) for the above four types can be arranged
such that one has a four-component Dirac equation at hand. This truly
remarkable mapping has led to the predictions that a number of Particle
Physics phenomena, such as Klein tunnelling, Andreev reflection among
others, can be (and have been) observed in graphene \cite{rev}.

However, from the present literature on graphene, it can be generally
observed that explicit computations in the context of graphene are not
performed in the four component Dirac form. In the present paper we provide
a new approach where explicit use has been made of the Dirac framework. We
exploit a generalized Foldy-Wuthouysen (FW) formalism \cite{fw} used for
four component Dirac Hamiltonian (the reason to use the four component Dirac
form is thus mainly technical), recently developed by some of us \cite%
{ber0,ber1,ber2}, where one can study the Dirac Hamiltonian with very
general type of interaction. The result comes as a series in powers of $%
\hbar $. It is very important to stress that in the present work, we have
also formulated a method of projection that can isolate the \textit{single}
valley effects (\textit{via } two component Pauli spinors) from the Dirac
spinor expressions and hence our method can be used to study single valley
and inter-valley interaction effects.

The method relies heavily on the appearance of \textit{intrinsic }Berry
gauge potentials \cite{berry,niu1} that induce a non-commutative geometry
(Berry curvature) in the electron phase space which subsequently affects the
particle dynamics \cite{BLIOKHDIRAC} (see also \cite{bl} for a review). In
the FW formalism, in order to properly identify the single particle
operators (see for example \cite{gr}), one needs to perform the same unitary
transformation $U$ on all the operators (position $\mathbf{r}$, momentum $%
\mathbf{p}$ etc.), that (partially) diagonalizes the Hamiltonian $H$, 
\begin{equation}
UHU^{+}=E~;~~\mathbf{r}=U\mathbf{r}U^{+}=\mathbf{R}+i\hbar U\mathbf{\nabla }%
U^{+}\equiv \mathbf{R}+\hbar \mathcal{A}^{R}  \label{H}
\end{equation}
Here $E$ denotes the diagonal energy eigenvalue matrix and $\mathcal{A}^{R}$
is the Berry potential. One can physically motivate the introduction of $%
\mathbf{r}$ by noting that in a Hamiltonian framework it yields the correct
expression for the velocity. Incidentally the configuration space $\mathbf{r}
$ (as well as the full phase space) turns out to be noncommutative. (The
earliest example of a noncommutative space (or phase space) was provided in 
\cite{sn}.) This is the ''algebraic'' reason for the emergence of Berry
phase.

On the other hand, intuitively the appearance of Berry phase effects is very
natural in FW formalism. The Berry potential is induced by the electron spin
coordinates which are treated as ''fast'' variables as compared to the
position coordinates which are considered as ''slow'' degrees of freedom.
The same phenomena was discovered earlier \cite{mead} when molecular
dynamics was studied in the Born-Oppenheimer approximation.

The advantage of our method is that one can study all these interactions in
a unified framework. In fact, for a constant distortion and for zero
magnetic field we recover the results of \cite{niu2}. But obviously more
interesting is the new results that we obtain for non-constant lattice
distortion, which is a new valley-orbit coupling due to the spatial
variation of the mass gap induced by the lattice distortion. This coupling
is similar to the valley-orbit coupling in the presence of an external
electric field and in turn contributes to the valley-Hall effect first
predicted in \cite{niu2}.

In this paper we also consider the semiclassical quantization of the
cyclotron orbit for an electron in a constant magnetic field and for a
constant mass. In particular we show the important role played by the valley
magnetic moment in the determination of the Landau levels.

The paper is organized as follows: In section \textbf{2} we discuss the
technicalities, construct the (non-commutative) covariant position and
momentum operators, identify the Berry potentials and compute the
semi-classical energy spectrum, in a generic framework. In Section \textbf{3}
we apply the scheme to graphene. In Section \textbf{3} we focus on effects
that result from the variable lattice distortion. Section \textbf{4} deals
with effects due to external electromagnetic fields. Section \textbf{5} is
devoted to discussions.

\section{Covariant phase space and Berry potentials}

We will explain our formalism through the concrete example of graphene.
Around the $K_{+}$ and $K_{-}$ points the energy spectrum is linear and the
dispersion relation can be approximated as $E_{\pm }\left( K\right) =\hbar
v_{F}\left| K-K_{\pm }\right| $. In the following $v_{F}$ is set to unity so
that the effective Hamiltonian near both valleys in the presence of a
distorted lattice, induced by a real bosonic field $\Delta \left( \mathbf{R}%
\right) $ and in the presence of an external electric potential $V\left( 
\mathbf{R}\right) $ can be written as 
\begin{equation}
H\left( \mathbf{P},\mathbf{R}\right) =\mathbf{\alpha .P}+\beta \Delta \left( 
\mathbf{R}\right) -eV\left( \mathbf{R}\right) .  \label{H1}
\end{equation}
where $e>0$ so that the charge of the electron is $-e$. In this expression
the energy is written in terms of the momentum\textbf{\ }$\mathbf{P}=\hbar 
\mathbf{K}$\textbf{\ }in order that\textbf{\ }$\hbar $\textbf{\ }appears
explicitly during the semiclassical diagonalization procedure described
below. We have\ also used the following representation for the Dirac
matrices 
\begin{equation*}
\alpha _{x}\mathbf{=}\left( 
\begin{array}{cc}
\sigma _{x} & 0 \\ 
0 & -\sigma _{x}%
\end{array}
\right) \text{ \ \ \ \ \ \ \ }\alpha _{y}\mathbf{=}\left( 
\begin{array}{cc}
\sigma _{y} & 0 \\ 
0 & \sigma _{y}%
\end{array}
\right) \text{ \ \ \ \ \ \ \ \ }\beta =\left( 
\begin{array}{cc}
\sigma _{z} & 0 \\ 
0 & \sigma _{z}%
\end{array}
\right)
\end{equation*}
where $\sigma _{i}$ are Pauli matrices. The dynamical operators $\mathbf{P}$
and $\mathbf{R}$ have components satisfying the canonical rule $\left[
R_{i},P_{j}\right] =i\hbar \delta _{ij}.$ We keep the $3+1$-dimensional
representation with the understanding that the components in the $z$%
-direction are null $R_{z}=P_{z}=0$ because graphene is a planar $2$%
-dimensional system.

Since we are interested in obtaining a semi-classical solution, we first
diagonalize the \textquotedblright classical\textquotedblright\ Hamiltonian $%
\left( \ref{H}\right) $ (whose arguments are not operators but classical
commuting variables noted $\left( \widetilde{\mathbf{P}}\mathbf{,}\widetilde{%
\mathbf{R}}\right) $) by a formal FW unitary transformation $U_{0}$ and
later reinsert the operators back. The $U_{0}$ in question is a conventional
FW matrix, 
\begin{equation}
U_{0}\left( \widetilde{\mathbf{P}}\mathbf{,}\widetilde{\mathbf{R}}\right) =%
\frac{E+\Delta \left( \widetilde{\mathbf{R}}\right) +\beta \mathbf{\alpha .}%
\widetilde{\mathbf{P}}}{\sqrt{2E(E+\Delta \left( \widetilde{\mathbf{R}}%
\right) )}}
\end{equation}%
such that the diagonal form of Hamiltonian is, 
\begin{equation}
U_{0}HU_{0}^{+}=\varepsilon _{0}\left( \widetilde{\mathbf{P}}\mathbf{,}%
\widetilde{\mathbf{R}}\right) =\beta \sqrt{\widetilde{\mathbf{P}}^{2}+\Delta
\left( \widetilde{\mathbf{R}}\right) ^{2}}-eV\left( \widetilde{\mathbf{R}}%
\right) .  \label{eq}
\end{equation}%
where $E=\sqrt{\widetilde{\mathbf{P}}^{2}+\Delta \left( \widetilde{\mathbf{R}%
}\right) ^{2}}$. After this \textquotedblright classical
diagonalization\textquotedblright\ the reintroduction of the quantum
operators $\mathbf{P}$ and $\mathbf{R}$ leads to the following semiclassical
diagonal Hamiltonian $H_{D}$ \cite{ber2}, 
\begin{equation}
\varepsilon \left( \mathbf{p,r}\right) \simeq \varepsilon _{0}\left( \mathbf{%
p,r}\right) +\frac{i\hbar }{2}\mathcal{P}_{+}\left[ \left[ \varepsilon
_{0}\left( \mathbf{p,r}\right) ,\mathcal{A}^{R_{l}}\right] \mathcal{A}%
^{P_{l}}-\left[ \varepsilon _{0}\left( \mathbf{p,r}\right) ,\mathcal{A}%
^{P_{l}}\right] \mathcal{A}^{R_{l}}\right] +O(\hbar ^{2}).  \label{epsilon}
\end{equation}%
In the above $\mathcal{P}_{+}$ denotes projection on the diagonal subspace
and 
\begin{equation}
\varepsilon _{0}\left( \mathbf{p,r}\right) =\beta \sqrt{\mathbf{p}%
^{2}+\Delta \left( \mathbf{r}\right) ^{2}}-eV\left( \mathbf{r}\right) .
\end{equation}%
In this expression the canonical operators $\left( \mathbf{R,P}\right) $
have been replaced by the covariant ones, 
\begin{eqnarray}
\mathbf{r} &=&\mathbf{R}+\hbar \emph{A}^{\mathbf{R}}  \notag \\
\mathbf{p} &=&\mathbf{P}+\hbar \emph{A}^{\mathbf{P}}  \label{rp}
\end{eqnarray}%
where we have also introduced the so called Berry connections 
\begin{equation*}
\emph{A}^{\mathbf{R}}=\mathcal{P}_{+}\left[ \mathcal{A}^{\mathbf{R}}\right]
~~;~~\emph{A}^{\mathbf{P}}=\mathcal{P}_{+}\left[ \mathcal{A}^{\mathbf{P}}%
\right] ,
\end{equation*}%
defined as the projection on the diagonal of $\mathcal{A}^{\mathbf{R}}=i%
\left[ U_{0}\nabla _{\mathbf{P}}U_{0}^{+}\right] $, $\mathcal{A}^{\mathbf{P}%
}=-i\left[ U_{0}\nabla _{\mathbf{R}}U_{0}^{+}\right] $. Here $U_{0}$ denotes
the matrix $U_{0}\left( \mathbf{P},\mathbf{R}\right) $ that is the matrix $%
U_{0}$ where now parameters have been replaced by operators.

Notice the reason for projecting on the diagonal is same as in the
conventional FW case \cite{gr} where this ensures that positive energy
solutions do not get mixed up with negative energy solutions. In the present
Condensed Matter scenario, these correspond to conduction band and valence
band.

From $U_{0}$ we can now deduce the following covariant coordinate and
momentum 
\begin{equation}
\mathbf{r}=\mathbf{R}-\hbar \frac{\mathbf{e}_{z}\times \mathbf{P}}{%
2E(E+\Delta )}\Sigma _{z},\text{ \ \ \ }\mathbf{p}=\mathbf{P,}  \label{r}
\end{equation}
where $\mathbf{e}_{z}$ is the unit vector in the $z-$direction and $\Sigma
_{z}=\left( 
\begin{array}{cc}
\sigma _{z} & 0 \\ 
0 & -\sigma _{z}%
\end{array}
\right) $. Notice that the momenta remain unchanged. This will be changed in
the presence of an external magnetic field.

From the general expression (\ref{epsilon}) we compute the full diagonal
quantum Hamiltonian, 
\begin{equation}
\varepsilon =\beta E-\frac{\hbar }{2E^{2}}\mathbf{e}_{z}\left( \mathbf{p}%
\times \mathbf{\nabla }_{\mathbf{r}}\Delta \left( \mathbf{r}\right) \right)
\Sigma _{z}-eV\left( \mathbf{r}\right) +O(\hbar ^{2})  \label{fullenergy}
\end{equation}
where now $E=\sqrt{\mathbf{p}^{2}+\Delta \left( \mathbf{r}\right) ^{2}}$.
However, all this is in the $3+1$-dimensional framework and to be applicable
to graphene we now need to reduce the model dimensionally. This we do in the
next section. The coordinate and momentum operators $\mathbf{r}$ and $%
\mathbf{p}$ are covariant operators because the energy operator (\ref%
{fullenergy}) expressed in terms of $\mathbf{r}$ and $\mathbf{p}$ is gauge
independent (i.e. independent of Berry connection).

\section{Single valley Hamiltonian for graphene}

In case of graphene one generally considers the two valleys to be isolated
and studies the electron behavior near one valley. This is achieved by
keeping the components ($1,2$) of the Hamiltonian for one valley and $\left(
3,4\right) $ for the second valley. Therefore the $2-$bands valley energy
can be indexed by $\tau =\pm 1$ such that : 
\begin{equation}
\varepsilon _{\tau }=\sigma _{z}\left( E-\tau \frac{\hbar }{2E^{2}}\mathbf{e}%
_{z}\left( \mathbf{p}\times \mathbf{\nabla }_{\mathbf{r}}\Delta \left( 
\mathbf{r}\right) \right) \right) -eV\left( \mathbf{r}\right)  \label{er}
\end{equation}%
where the eigenvalues $\pm 1$ of $\sigma _{z}$ corresponds to the conduction
and valence band energies of the valley $\tau $. The position operator also
is now given by the projection of (\ref{r}) on the components ($1,2$), 
\begin{equation}
\mathbf{r}=\mathbf{R}-\tau \frac{\hbar \mathbf{e}_{z}\times \mathbf{P}}{%
2E(E+\Delta )}\sigma _{z}  \label{roperator}
\end{equation}%
Now and for the rest of the paper we consider the properties of an electron
in a conduction band (the positive energy band of (\ref{er})) of a valley,
which means that the position operator also has to be projected on this
subspace leading to the expression 
\begin{equation}
\mathbf{r}=\mathbf{R}-\tau \frac{\hbar \mathbf{e}_{z}\times \mathbf{P}}{%
2E(E+\Delta )}
\end{equation}%
From this expression of the valley dependent coordinate operator we deduce
that the Algebra satisfied by the covariant dynamical variables is a non
commutative one expressed in terms of Berry curvatures $\Theta _{ij}^{\zeta
\eta }=\partial _{\zeta ^{i}}\emph{A}_{j}^{\eta }-\partial _{\eta ^{i}}\emph{%
A}_{j}^{\zeta }$ where $\zeta ,\eta $ mean either $r$ or $p$, 
\begin{eqnarray}
\left[ x,y\right] &=&i\hbar ^{2}\Theta _{xy}^{rr}=-i\hbar ^{2}\frac{\tau
\Delta }{2E^{3}}  \label{commut} \\
\left[ p_{i},r_{j}\right] &=&-i\hbar \delta _{ij}+i\hbar ^{2}\Theta
_{ij}^{pr}=-i\hbar \delta _{ij}+\tau \frac{i\hbar ^{2}}{2E^{3}}\varepsilon
^{jk3}P_{k}\mathbf{\nabla }_{R_{i}}\Delta  \label{xp} \\
\left[ p_{x},p_{y}\right] &=&i\hbar ^{2}\Theta _{xy}^{pp}=0\text{ }
\label{pp}
\end{eqnarray}%
The Berry curvature $\Theta _{xy}^{rr}=-\frac{\tau \Delta }{2E^{3}}$
obtained here is denoted in the rest of the paper simply by $\Theta $. We
mention that the general expression in (\ref{commut}) reduces to that of 
\cite{niu2} who considered valley Hall effect for a \textit{constant}
distortion $\Delta $. We, on the other hand, show that for the more general
case of non-constant distortion, there are additional Berry curvature
components of the mixed form $\Theta _{ij}^{pr}$. In the absence of a
magnetic field the Berry curvature $\Theta _{xy}^{pp}$ is zero.
interestingly, the energy can be rewritten in terms of the Berry curvature
as $\varepsilon _{\tau }=E-eV\left( \mathbf{r}\right) +\hbar E\left( \Theta
_{xx}^{pr}+\Theta _{yy}^{pr}\right) .$

At this level, it is interesting to express the conduction band energy in
terms of the canonical coordinates $\mathbf{P}$ and $\mathbf{R}$ which reads 
\begin{equation}
\varepsilon _{\tau }=\sqrt{\mathbf{P}^{2}+\Delta \left( \mathbf{R}\right)
^{2}}-\frac{\hbar \tau }{2E\left( \Delta +E\right) }\mathbf{P}\times \left[
\left( \frac{2\Delta +E}{E}\right) \mathbf{\nabla }_{\mathbf{R}}\Delta -e%
\mathbf{\nabla }_{\mathbf{R}}V\right] \cdot \mathbf{e}_{z}-eV(\mathbf{R})
\label{1}
\end{equation}%
For a constant mass term $\Delta =cst$, we observe a valley-orbit coupling
term $\frac{\hbar \tau }{2E\left( \Delta +E\right) }\mathbf{P}\times e%
\mathbf{\nabla }_{\mathbf{R}}V$ which is of the same origin as that of the
fully relativistic spin orbit coupling $\frac{\hbar }{2E\left( m+E\right) }%
\sigma .(\mathbf{P}\times e\mathbf{\nabla }_{\mathbf{R}}V)$ in the Dirac
equation \cite{ber0}\cite{NIU3}. In the literature this coupling is usually
written in the non-relativistic limit only $\frac{\hbar }{4m^{2}}\sigma .(%
\mathbf{P}\times e\mathbf{\nabla }_{\mathbf{R}}V)$ because it comes out from
the Foldy-Wouthuysen transformation which is an expansion in $1/m.$ The
semiclassical diagonalization allows one to consider the fully relativistic
situation. Interestingly for a non constant mass term we observe an
additional valley-orbit coupling $\hbar \tau \left( \frac{2\Delta +E}{%
2E^{2}\left( \Delta +E\right) }\right) \mathbf{P}\times \mathbf{\nabla }_{%
\mathbf{R}}\Delta $. Therefore \textit{a variable mass term mimics an
external electric field} and can thus lead to similar effects as with a true
electric field as discuss in next section. All this shows that the
non-commutative position operator encodes through the Berry potential the
spin or valley-orbit coupling. In particular, as discussed in \cite{ber0},
the dynamics of the canonical operators turn out to be different from that
of the non-commutative ones.

\section{Effect of variable lattice distortion $\Delta (\mathbf{r})$}

As already stated, $\mathbf{\nabla }_{\mathbf{R}}{\Delta }$ term induces an
additional valley-orbit coupling. More precisely, the field $\Delta \left( 
\mathbf{R}\right) $ is responsible for a valley-orbit coupling term $\sim 
\mathbf{P}\times \mathbf{\nabla }_{\mathbf{R}}\Delta $ similar to the
valley-orbit coupling induced by the electric field $\sim \mathbf{P}\times 
\mathbf{\nabla }_{\mathbf{R}}V$ which in turn is of the same nature as the
spin-orbit coupling for Dirac electrons in the non-relativistic limit (Pauli
term). Our results reveal that \textit{\ a defect of the lattice distortion
can actually be interpreted as an effective electric field}, which can be
responsible of a mechanical force leading to a valley Hall effect as the
electric field does \cite{niu2}. This is closely related to the spin Hall
effect in semiconductors \cite{MURAKAMI},\cite{SINOVA}). This is a new
observation. This clearly reminds us of the work of Bernevig and Zhang \cite%
{bern} who showed that shear strain in zinc-blende semiconductors such as $%
GaAs$ induces a spin-orbit coupling term which mimics the usual spin-orbit
in the presence of an electric field (this mechanism in \cite{bern} then
leads to a quantum spin Hall effect, which is different from the effect
considered here ).

Interestingly within our approach we can study the effect of the
valley-orbit coupling in both \textquotedblright
ultra-relativistic\textquotedblright\ ($\Delta <<P$) and \textquotedblright
non-relativistic\textquotedblright\ limits ($\Delta >>P$).

In the regime of a large gap $\Delta >>P,$ the energy becomes 
\begin{equation}
\varepsilon _{\tau }\approx \Delta \left( \mathbf{R}\right) +\frac{\mathbf{P}%
^{2}}{2\Delta \left( \mathbf{R}\right) }-\frac{\hbar \tau }{4\Delta \left( 
\mathbf{R}\right) ^{2}}\mathbf{P}\times \left[ 3\mathbf{\nabla }_{\mathbf{R}%
}\Delta (\mathbf{R})-e\mathbf{\nabla }_{\mathbf{R}}V(\mathbf{R})\right]
\cdot \mathbf{e}_{z}  \label{2}
\end{equation}%
which shows the same behavior than the usual non-relativistic limit energy
of the Dirac Hamiltonian. For a constant $\Delta $, the valley-orbit
contribution is cancelled by the presence of the mass-gap term $4\Delta ^{2}$
which is large compared to electron momentum (this is similar to the mass
term $4m^{2}c^{2}$ in the Pauli term of non-relativistic Dirac electron).
But this coupling can be greatly enhanced in the presence of a very sharp
variation of the lattice distortion $\Delta $. This effect could be much
larger than the valley-orbit coupling in the presence of an electric field
if $\mathbf{\nabla }_{\mathbf{R}}\Delta >>e\mathbf{\nabla }_{\mathbf{R}}V.$
Such very localized lattice distortion defect might be built by adsorbing
the graphene on two different substrates in contact creating in this way a
line defect\textbf{.}

In the opposite regime $\Delta <<P$ we have 
\begin{equation}
\varepsilon _{\tau }=P-\frac{\hbar \tau }{2P^{2}}\mathbf{P}\times \left[ 
\mathbf{\nabla }_{\mathbf{R}}\Delta -e\mathbf{\nabla }_{\mathbf{R}}V\right]
\cdot \mathbf{e}_{z}  \label{3}
\end{equation}%
Here the valley-orbit coupling is independent of the magnitude of the
distortion field and can also be greatly enhanced by a very sharp variation
of the distortion $\Delta $.

Therefore in both cases, a large gradient of the lattice distortion will
generate a bigger valley-orbit interaction than with the presence of an
electric field. This coupling may in principle induce a valley-Hall effect
very similar to the valley-Hall effect considered in \cite{niu2}, the latter
having an electric field and a uniform $\Delta $ .

This valley-Hall effect is better seen in the equations of motion which are
given in terms of the covariant operators. The above noncommutative algebra (%
\ref{commut}) leads to a non-trivial modification in the dynamics of an
electron in a conduction band of a valley which can be written 
\begin{equation}
\mathbf{\dot{r}}=\mathbf{\nabla }_{\mathbf{p}}\varepsilon \left( 1-\hbar
\Theta ^{pr}\right) -\hbar \mathbf{\dot{p}\times \Theta },\text{ \ \ \ }%
\mathbf{\dot{p}}=-\mathbf{\nabla }_{\mathbf{r}}\varepsilon \left( 1-\hbar
\Theta ^{pr}\right)  \label{4}
\end{equation}%
where we have defined the curvature vector $\mathbf{\Theta }=\Theta \mathbf{e%
}_{z}$ and $\mathbf{\nabla }_{\mathbf{p}}\varepsilon \left( 1-\hbar \Theta
^{pr}\right) $ has the following meaning $\nabla _{p^{j}}\varepsilon \left(
\delta _{ij}-\hbar \Theta _{ij}^{pr}\right) $.

In more details at the first order in $\hbar $ we obtain for the momentum
dynamics

\begin{equation}
\mathbf{\dot{p}=-}\frac{\Delta }{E}\nabla _{\mathbf{r}}\Delta +e\nabla _{%
\mathbf{r}}V  \label{5}
\end{equation}
such that the velocity reads:

\begin{equation}
\mathbf{\dot{r}}=\frac{\mathbf{p}}{E}-\frac{\hbar \tau }{2E^{3}}\left(
\left( \frac{p^{2}+2\Delta ^{2}}{E}\right) \mathbf{\nabla }_{\mathbf{r}%
}\Delta -\Delta e\mathbf{\nabla }_{\mathbf{r}}V\right) \times \mathbf{e}_{z}
\label{rrrr}
\end{equation}%
Importantly, the lattice distortion $\Delta \left( \mathbf{R}\right) $
induces an anomalous velocity $\mathbf{\dot{p}\times \Theta }$ even in the
absence of an electric field. Suppose that the distortion is in the $y-$
direction. Then we get a spontaneous (non-steady) current both in the
longitudinal ($y-$direction) and in the transverse $x-$direction due to the
anomalous velocity. Then electrons will accumulate irrespective of their
magnetic moment at the longitudinal edge. And a electric potential between
the edges in the $y-$ direction can be measured. In the same manner, we will
have a spontaneous separation of the electrons depending on the value of
their magnetic moments and thus an accumulation in the transverse edges. A
steady currents requires an electric field. If the mass-valley-orbit
coupling is stronger than the electric-valley-orbit coupling, then the
dominant transverse current is due to the lattice distortion and leads a
transverse velocity $v_{x}\simeq \frac{-\hbar \tau }{2E^{3}}\left( \frac{%
p^{2}+2\Delta ^{2}}{E}\right) \nabla _{y}\Delta $ which depends of the
valley index but is charge independent. Therefore there will be a transverse
current in the $x-$direction $j_{x}\left( \mathbf{r}\right) \simeq e\frac{%
\hbar \tau }{2}\int d\mathbf{p}~g\left( \mathbf{p},\mathbf{r}\right) \left( 
\frac{p^{2}+2\Delta ^{2}}{E^{4}}\right) \mathbf{\nabla }_{y}\Delta $ where $%
g\left( p,r\right) $ is the statistical distribution of carrier, leading to
a Valley Hall effect (separation of the carrier depending of the valley).
Here we considered the electron in a ballistic regime neglecting scattering
with the impurities. If the the system is not connected to the leads along
the $x-$direction, carriers of different valleys accumulate near the edges
of the sample as discussed in \cite{niu2}. If the density of up and down
magnetic moments are equal there will be not potential difference in the $x-$%
direction. To observe a transverse electric potential one needs to control
the chemical potential of both species as suggested in \cite{niu2}.

We now consider the case of the graphene in an external electromagnetic
field.

\section{Graphene in a magnetic field}

The magnetic field interaction is introduced via minimal coupling and the
Dirac Hamiltonian becomes, 
\begin{equation}
H\left( \mathbf{P},\mathbf{R}\right) =\mathbf{\alpha .\Pi }+\beta \Delta
\left( \mathbf{R}\right) +V\left( \mathbf{R}\right)  \label{61}
\end{equation}%
with $\mathbf{\Pi =\mathbf{P+}}e\mathbf{\mathbf{A}}$ the covariant momentum.
The semiclassical evolution of a Dirac electron in an electromagnetic field
for a constant mass term was first derived in reference \cite{BLIOKHDIRAC}.
For the generalization to a variable mass we use the general method of
Hamiltonian diagonalization of \cite{ber2}.

\subsection{Hamiltonian diagonalization}

As shown in \cite{ber2}, the introduction of the gauge potential is a
straightforward generalization of Eq. (\ref{er}) and one obtains for the
valley $\tau $ the following diagonal representation of the valley
Hamiltonian for a carrier in the conduction band

\begin{equation}
\varepsilon _{\tau }=\sqrt{\mathbf{\pi }^{2}+\Delta \left( \mathbf{r}\right)
^{2}}-\tau \frac{\hbar }{2E^{2}}\mathbf{e}_{z}\left( \mathbf{\pi }\times 
\mathbf{\nabla }\Delta \left( \mathbf{r}\right) \right) +e\hbar E\Theta
B_{z}-eV\left( \mathbf{r}\right)  \label{EnergyB}
\end{equation}%
with $\mathbf{r}=\mathbf{R}+\hbar \emph{A}^{\mathbf{R}}$, $\mathbf{\pi }=%
\mathbf{\Pi -}e\hbar \emph{A}^{\mathbf{R}}\times \mathbf{B}$ and $\mathbf{B}=%
\mathbf{\nabla }\times \mathbf{A.}$ The covariant dynamical operators are
explicitly given by 
\begin{equation}
\mathbf{r}=\mathbf{R}-\tau \frac{\hbar \mathbf{e}_{z}\times \mathbf{\Pi }}{%
2E(E+\Delta )}\text{ \ and \ }\mathbf{\pi }=\mathbf{\Pi }+\tau \frac{e\hbar 
\mathbf{\Pi }}{2E(E+\Delta )}B_{z}  \label{Rpi}
\end{equation}%
where $E=\sqrt{\mathbf{\pi }^{2}+\Delta \left( \mathbf{r}\right) ^{2}}.$ We
note that in the presence of a magnetic field the momentum get a Berry
connection a well. Therefore the new commutation relations are, 
\begin{eqnarray}
\left[ r_{x},r_{y}\right] &=&i\hbar ^{2}\Theta =-i\hbar ^{2}\frac{\tau
\Delta }{2E^{3}} \\
\left[ \pi _{i},r_{j}\right] &=&-i\hbar \delta _{ij}-ie\hbar ^{2}\Theta
_{il}\varepsilon ^{lj3}B_{z}-i\hbar ^{2}\frac{\tau e}{2E^{3}}\varepsilon
^{jk3}\mathbf{\Pi }_{k}\mathbf{\nabla }_{R_{i}}\Delta
\end{eqnarray}%
and we obtain in addition the non trivial commutation relation between the
covariant momenta, 
\begin{equation}
\left[ \pi _{x},\pi _{y}\right] =-ie\hbar B_{z}+ie^{2}\hbar ^{2}\Theta
B_{z}^{2}.  \label{ppp}
\end{equation}%
We notice also that only the component of the magnetic field in the $z$%
-direction (perpendicular to the graphene plane) appears in the energy and
in the definition of the covariant momentum. This peculiarities is due to
the $2$-dimensional nature of the graphene.

From the above expressions it is also natural to introduce the magnetization
vector (similar to the spin) 
\begin{equation}
m\left( \mathbf{\pi }\right) =-e\hbar E\Theta =\frac{e\hbar \tau \Delta }{%
2E^{2}}
\end{equation}%
so that interaction with the magnetic field reads $-mB_{z}$ with the
magnetic moment $m$ being an intrinsic one, associated to a valley and
originated from Berry curvature. The magnetic moment which interacts with
the magnetic field in the $z-$direction only is an important quantity which
allows to make a distinction between the carriers of the two valleys as
proposed by Xiao et.al. \cite{niu2}. Here we will show that the
magnetization is an essential quantity for the computation of the energy
levels (Landau levels).

\subsection{Semiclassical quantization of cyclotron orbit}

Thus to simplify the discussion consider a zero electric field $V\left( 
\mathbf{r}\right) =0$, a constant lattice distortion $\Delta \left( \mathbf{R%
}\right) =cst=\Delta $ and a constant magnetic field $\mathbf{B=}B\mathbf{e}%
_{z}$.

We start with the general semiclassical equations of motion for electrons in
magnetic Bloch bands \ 
\begin{equation}
\mathbf{\dot{r}}=\frac{\partial \varepsilon }{\partial \mathbf{\pi }}-\hbar 
\overset{\cdot }{\mathbf{\pi }}\times \mathbf{\Theta },\text{ and \ }\overset%
{\cdot }{\mathbf{\pi }}=-e\mathbf{\dot{r}\times B}
\end{equation}
where $\varepsilon $ is the energy of a magnetic Bloch band. We have an
anomalous velocity contribution $\overset{\cdot }{\mathbf{\pi }}\times 
\mathbf{\Theta }$ which is the ''dual'' of the relativistic Lorentz force $%
\mathbf{\dot{r}\times B}$ (for a non-uniform $B$ one must add to the second
equation a kind of Stern and Gerlach term $-m\mathbf{\nabla }B$ which
couples the magnetization to the gradient of the magnetic field).

Combining both equations we have 
\begin{equation}
\mathbf{\dot{r}}=\frac{1}{1-e\hbar \mathbf{B.\Theta }}\left( \frac{\partial
\varepsilon _{\tau }}{\partial \mathbf{\pi }}\right) \text{ \ and \ }\overset%
{\cdot }{\mathbf{\pi }}=\frac{-e}{\left( 1-e\hbar \mathbf{B.\Theta }\right) }%
\left( \frac{\partial \varepsilon _{\tau }}{\partial \mathbf{\pi }}\mathbf{%
\times B}\right)
\end{equation}
These are the usual semiclassical equations of motion of an electron in a
magnetic Bloch band except for the presence of the term $1-e\hbar \mathbf{%
B.\Theta }$ which as $\mathbf{B=}B\mathbf{e}_{z}$ and $\mathbf{\Theta }%
=\Theta \mathbf{e}_{z}$ becomes $1-e\hbar B\Theta .$ For the graphene case
we have $\varepsilon =\varepsilon _{\tau }=\sqrt{\mathbf{\pi }^{2}+\Delta
^{2}}-mB$. Therefore, the equations of motion to the first order in $\hbar $
becomes 
\begin{equation}
\mathbf{\dot{r}}=\mathbf{\pi }\left( \frac{1-e\hbar B\Theta }{E}\right) ,\ \
\ \ \ \overset{\cdot }{\mathbf{\pi }}=e\mathbf{\pi \times B}\left( \frac{%
1-e\hbar B\Theta }{E}\right)  \label{EQmotionB}
\end{equation}
which are the same than for a relativistic particle in a magnetic field
except for the additional multiplicative factor $1-e\hbar B\Theta .$ Note
that we have $\mathbf{\pi }\frac{d\mathbf{\pi }}{dt}=0$ so that $E$ is a
constant of motion, and also $\frac{d\Theta }{dt}=0$ and $\frac{dm}{dt}=0.$

We now focus on the non-relativistic case $\Delta ^{2}>>\mathbf{\pi }^{2}$
so that $\varepsilon _{\tau }\approx \frac{\mathbf{\pi }^{2}}{2\Delta ^{2}}%
+\Delta -mB$ and in terms of the coordinates Eq. (\ref{EQmotionB}) reads 
\begin{equation}
\overset{\cdot }{x}=\pi _{x}\left( \frac{1-e\hbar B\Theta }{\Delta }\right) 
\text{, \ \ \ }\overset{\cdot }{y}=\pi _{y}\left( \frac{1-e\hbar B\Theta }{%
\Delta }\right)  \label{12}
\end{equation}%
and 
\begin{equation}
\overset{\cdot }{\pi }_{x}=-e\pi _{y}B\left( \frac{1-eB\hbar \Theta }{\Delta 
}\right) ,\text{ \ \ \ \ }\overset{\cdot }{\pi }_{y}=e\pi _{x}B\left( \frac{%
1-eB\hbar \Theta }{\Delta }\right)  \label{11}
\end{equation}%
Choosing the gauge $A_{x}=0$ and $A_{y}=BX=B(x-\tau \frac{\hbar \mathbf{\pi }%
_{y}}{4\Delta ^{2}})$ we have $\pi _{y}=p_{y}+eB(x-\tau \frac{\hbar }{%
4\Delta ^{2}}\pi _{y})+\tau \frac{e\hbar \pi _{y}}{4\Delta ^{2}}B=p_{y}+eBx$
(note the importance of the Berry connection here), which gives $\overset{%
\cdot }{p}_{y}=\overset{\cdot }{\pi }_{y}-eB\overset{\cdot }{x}=0$ thus $%
p_{y}$ is a constant of motion. Or as $\left( 1-e\hbar B\Theta \right)
/\Delta $ is a constant of motion we can write 
\begin{equation}
\overset{\cdot \cdot }{x}=\omega ^{2}\left( x_{0}-x\right)  \label{10}
\end{equation}%
which is the equation of a harmonic oscillator with the orbit center $x_{0}=%
\frac{p_{y}}{-eB}$ which is the usual expression except that it is expressed
in terms of the covariant coordinate and momentum and not in terms of the
canonical ones. The cyclotron frequency $\omega =\frac{Be}{\Delta }\left(
1-e\hbar B\Theta \right) $ is also the usual result for the quantization of
a non relativistic electron in a magnetic field except that it is slightly
corrected. But this correction is actually negligible for the computation of
the energy levels and we have 
\begin{equation}
E_{n}\approx \pm \left( \frac{\hbar eB}{\Delta }\left( n+\frac{1}{2}-\frac{%
\tau }{2}\right) +\Delta \right) +O\left( \hbar ^{2}\right)
\end{equation}%
where the contribution $\frac{\tau }{2}$ comes from the magnetization term.
At the semiclassical order this expression can also be written 
\begin{equation}
E_{n}\approx \pm \sqrt{\Delta ^{2}+2\hbar Be\left( n+\frac{1}{2}-\frac{\tau 
}{2}\right) }+O\left( \hbar ^{2}\right)  \label{EN}
\end{equation}%
Therefore the ground state is not degenerate as there is only one
possibility to realize it $n=0$ and magnetization up. We mention here the
lattice distortion in the present case that leads to a mass in the Dirac
equation is imposed by hand. In a very interesting paper \cite{FUCHSPRL}, an
explicit mechanism relying on an electron-lattice interaction in the
presence of an external magnetic field is proposed for the lattice
distortion and which leads to the energy levels Eq. (\ref{EN}).

It is interesting to show that one can easily recover this result from the
semiclassical Bohr-Sommerfeld quantification rule 
\begin{equation}
\oint P_{x}dX=2\pi \hbar \left( n+1/2\right)  \label{BS}
\end{equation}%
From (\ref{Rpi}), $\mathbf{\pi }=\mathbf{\mathbf{P+}}e\mathbf{\mathbf{A}}%
+\tau \frac{e\hbar \mathbf{\Pi }}{2E(E+\Delta )}B$, we deduce that $%
P_{x}=\pi _{x}-\tau \frac{e\hbar \pi _{x}B}{2E(E+\Delta )}$ (for $A_{x}=0,$
and $\mathbf{\Pi }_{x}$ can safely replace by $\pi _{x}$ at this order in $%
\hbar $). In the same manner (\ref{Rpi}) gives $dX=dx-\tau \frac{\hbar d%
\mathbf{\pi }_{y}}{2E(E+\Delta )}$ (as $E$ is a constant of motion). This
leads to 
\begin{equation}
\mathbf{\pi }^{2}\left( 1-\tau \frac{e\hbar B}{E(E+\Delta )}\right) =2\hbar
eB\left( n+1/2\right) ,  \label{pin}
\end{equation}%
where $\pi ^2=\pi_x^2 +\pi _y^2$. In the case $\Delta ^{2}>>\mathbf{\pi }^{2}
$, we can replace $\mathbf{\pi }^{2}$ by $\left( E_{n}+mB\right) ^{2}-\Delta
^{2}$ and neglect the contribution $\tau \frac{e\hbar B}{E(E+\Delta )},$ so
that 
\begin{equation}
E_{n}\approx \sqrt{\Delta ^{2}+2\hbar Be\left( n+\frac{1}{2}-\frac{\tau }{2}%
\right) }
\end{equation}%
and we retrieve the previous result. Note a similar derivation based on the
Mikitik and Sharlai's theorem \cite{Mikitik} about the Berry phase in the
Onsager semiclassical quantization of cyclotron orbits is provided in \cite%
{Fuchs}.

In the opposite regime $\Delta ^{2}<<\mathbf{\pi }^{2},$ $\pi ^{2}\approx
\varepsilon _{\tau }^{2}$ and (\ref{pin}) therefore becomes $E_{n}^{2}-\tau
e\hbar B=2\hbar eB\left( n+1/2\right) $ so that the Landau levels are given
by 
\begin{equation}
E_{n}=\pm \sqrt{2\hbar eB\left( n+\frac{1}{2}-\frac{\tau }{2}\right) }
\end{equation}
Interestingly from this semiclassical analysis we retrieve the exact result.
It is important to stress that here the contribution $\frac{\tau }{2}$ does
not come from the magnetization term which is zero here but from the Berry
connection in momentum $\emph{A}^{\mathbf{P}}$ which in the non-relativistic
case turns out to be negligible.

\section{Conclusion}

The two-dimensional carbon crystalline honeycomb structure - graphene - has
created an enormous amount of activity because of its striking electronic
properties. Near each of the two distinct and well separated valleys (or
Dirac points) in the Brillouin zone, the energy dispersion of electrons has
a linear structure with electrons moving with an (energy independent) Fermi
velocity. It means that the electrons can be regarded as massless
relativistic two dimensional Dirac particles. Thus a number of exotic
relativistic phenomena are, in principle, observable in graphene. It is
possible to simultaneously consider the two valleys in a four dimensional
framework and this allows us to apply the machinery of four component Dirac
electron in the planar system of graphene.

In the present paper we have exploited a generalized form \cite{ber2} of the
well known Foldy Wouthuysen \cite{fw} transformation to generate the energy
spectrum (in a semi-classical quantization scheme). We construct the
covariant position and momentum operators which, on account of the intrinsic
Berry potentials, satisfy a non-commutative phase space algebra. The induced
Berry curvature terms directly affect the electron dynamics, as for example
by generating an anomalous velocity term.

The main advantages of employing the generalized Foldy Wouthuysen formalism 
\cite{ber2} is twofold: \newline
(i) Different forms of interactions can be studied in a \textit{unified}
way. \newline
(ii) The approximation scheme leads in a straightforward way to
semi-classical results which are thus valid for a small ratio of the
electron wavelength to the typical spatial scale of external perturbations
(either in mass or in external potential). \newline
Referring to (i) above, in the present paper we have discussed effects of an
arbitrary position dependent lattice distortion which gives rise to a novel
\textquotedblright valley-orbit\textquotedblright\ coupling (similar to the
spin-orbit coupling) where one observes that the lattice distortion induces
an effective electric field. Obviously our general expressions (such as
Berry curvature) reduce to the well known results of \cite{niu2} for
constant lattice defect. It is important to stress that in both cases the
intrinsic inhomogeneity of the lattice drives the effect without any
external electric field. We have also studied of effects external electric
and magnetic fields. \newline
In the context of (ii) above, we have shown that one can study both the
\textquotedblright ultra-relativistic\textquotedblright\ and
\textquotedblright non-relativistic\textquotedblright\ limits in a single
setup depending on the effective mass-gap (or lattice defect $\Delta $)
being smaller or larger than the electron momentum.

\begin{acknowledgement}
The authors acknowledge fruitful discussions with F. Pi\'{e}chon and J.N.
Fuchs. We thank the referees for the constructive comments.
\end{acknowledgement}

\end{document}